\documentclass[useAMS,usenatbib]{mn2e}
\usepackage{times}
\usepackage{graphicx}
\usepackage{epstopdf}

\title[RAT\,J0455+1305: A rare hybrid pulsating
subdwarf B star]{RAT\,J0455+1305: A rare hybrid pulsating
subdwarf B star}
\author[A. S. Baran et al.]{A. S. Baran,$^{1,2}$\thanks{E-mail:
asb@iastate.edu}\thanks{ASB and JTG: Visiting Astronomers, Kitt Peak National Observatory, National Optical Astronomy Observatory, which is operated by the Association of Universities for Research in Astronomy (AURA) under cooperative agreement with the National Science Foundation} J. T. Gilker,$^{3}$ L. Fox-Machado,$^{4}$ M. D. Reed$^{3}$ and S. D. Kawaler$^{2}$\\
$^{1}$ Cracow Pedagogical University, ul. Podchorazych 2, 30-084 Krakow, Poland\\
$^{2}$ Department of Physics and Astronomy, Iowa State University, Ames, IA 50011, USA\\
$^{3}$ Department of Physics, Astronomy, and Materials Science, Missouri State University, Springfield, MO 65897 USA\\
$^{4}$ Instituto de Astronom\'{\i}a Ð Universidad Nacional Aut\'onoma de M\'exico, Ap. P. 877, Ensenada, BC 22860, Mexico}

\begin{document}

\date{}

\pagerange{\pageref{firstpage}--\pageref{lastpage}} \pubyear{2010}

\maketitle

\label{firstpage}

\begin{abstract}
We present results on the second-faintest pulsating subdwarf B (sdB) star known, RAT\,J0455+1305, derived from photometric data obtained in 2009. It shows both short and long periods oscillations, theoretically assigned as pressure and gravity modes. We identify six short-period frequencies (with one being a combination) and six long-period frequencies. This star is the fourth hybrid sdB star discovered so far which makes it of special interest as each type of mode probes a different part of the star. This star is similar to the sdB hybrid pulsator Balloon\,090100001 in that it exhibits short-period mode groupings, which can be used to identify pulsation parameters and constrain theoretical models.

\end{abstract}


\section{Introduction}
Hot subdwarf (sdB) stars are horizontal branch stars with masses near to 0.5 M$_\odot$
and very thin (in mass) hydrogen envelopes. Average effective temperatures and surface gravities are about 
30\,000~K and $\log g \sim$\,5.5, respectively. Although it is clear that sdB stars will eventually enter the white dwarf cooling track without reaching the AGB phase \citep{dorman93}, their formation as sdB stars is under debate.  There are several mechanisms that involve single-star or binary evolution \citep[e.g.][]{dcruz96,han02,han03}.
Detection of pulsations in hot sdB stars opened a way to study their interiors and evolution prior to the 
horizontal branch. First, short period oscillations were found by \cite{kilkenny97} in EC\,14026 (now officially named 
V391\,Hya). According to theoretical models these pulsations are attributed to pressure ($p-$)modes 
and are driven in the outer 
part of the stars \citep{charp97}. Several years after their discovery, \cite{green03} announced another kind of 
variability in sdB stars. The longer-period variations were also identified as stellar oscillations. In this case however, 
they are attributed to gravity ($g-$)modes and originate deeper within these stars than the ($p-$)modes \citep{fontaine03}.  
Both types share some overlap in the H--R diagram and so the same stellar models are appropriate for studying both kinds 
of pulsation. Of special interest are these rare stars showing both ($p-$) and ($g-$)modes since models of these 
stars can constrain their cores and outer regions simultaneously. The first hybrid sdBV star was found by \cite{schuh06}
and three more have been discovered using ground-based data, with RAT\,J0455+1305 being the faintest. The Kepler satellite has also recently discovered several candidate hybrid sdBV stars \citep{reed10,kawaler10,ostensen10b} from its survey phase with follow-up data to be obtained over the next several years.

The brightest known pulsating subdwarf B (sdBV) star has a $V$ magnitude of 11.8, the typical brightness is 
about 14.5 and the faintest are 17 \citep{ostensen10a}. Methods for constraining the pulsation modes include
multicolor photometry \citep{tremblay06}, low-resolution spectroscopy \citep{telting04,telting06,reed09}, a combination of these
two \citep{baran08}, and time-series observations of line profile variations using high-resolution spectrosocpy \citep{telting08,telting10}. This last method requires large telescopes, where it is quite difficult to get time. To date, time-series spectroscopy has been obtained for only a few objects, all brighter than 14\,mag. Because of the difficulty of obtaining data spanning relatively long periods of time, the faintest stars are usually limited to discovery observations taken in one or no filter at all (so-called white light).

Single-color data, which are typically obtained to increase temporal resolution, are usually not useful for mode 
identification, which is necessary for constraining the models. To identify modes using broadband single color data 
one needs to assume trial values for free parameters in models and search for the best fit to observed frequencies. 
The large number of free parameters does not lead to strong constraints on stellar models. An exception is when the 
frequency spectrum contains specific features, such as frequency multiplets, mode groupings, or small and large 
spacings, which constrain mode identifications and therefore reduce the number of free parameters. 

Following this approach, and encouraged by the results obtained for other sdBV stars, we decided to obtain photometric 
observations of one of the faintest sdBV stars discovered so far, RAT\,J0455+1305 (hereafter RAT0455). Our results
clearly show that small and/or moderate telescopes are useful for the asteroseismic study of faint sdBV stars.

\begin{table}
 \centering
 \begin{minipage}{83mm}
  \caption{Log of observations}
  \label{log}
  \begin{tabular}{@{}ccccc@{}}
  \hline
   Date              &   hours   &   exposure   &   filter   &           site  \\
 \hline
10 Nov 2009  &     7.2      &         30s       &  none   & San Pedro Martir  \\
11 Nov 2009  &     7.4      &         30s       &  none   & San Pedro Martir  \\
14 Nov 2009  &     7.7      &         30s       &  none   & San Pedro Martir  \\
24 Dec 2009  &     1.9      &         30s       &  BG40  & KPNO  \\
25 Dec 2009  &     8.6      &         30s       &  BG40  & KPNO  \\
26 Dec 2009  &     8.6      &         30s       &  BG40  & KPNO  \\
27 Dec 2009  &     8.4      &         30s       &  BG40  & KPNO  \\
\hline
\end{tabular}
\end{minipage}
\end{table}

\section{RAT0455}
\begin{figure}
\includegraphics[width=83mm]{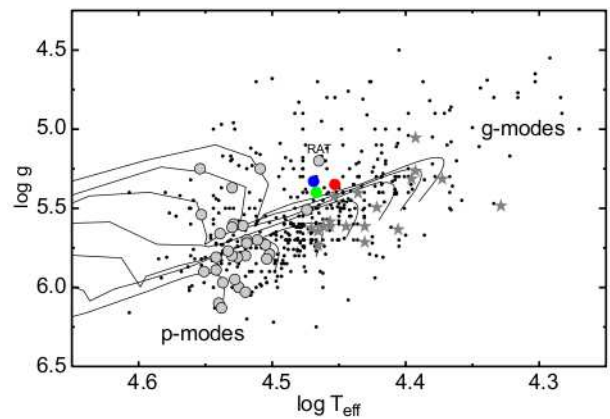}
\caption{Location of hybrid stars in the logT$_{\rm eff}$ -- $\log g$ plane. Previously known hybrid stars 
are marked with colors: Balloon\,090100001-blue, DW\,Lyn-red, V391\,Peg-green. The location of RAT0455 is marked with 'RAT'.}
\label{hr}
\end{figure}

RAT0455 was discovered during the RApid Temporal Survey \citep{ramsey05}.
The aim of this survey was to search for variability
from a few minutes up to several hours, for stars down to 22\,mag
to find interacting ultra-compact binary systems with orbital
periods less than 70\,min. The design of the project is also useful for detecting sdBV
stars, which can serendipitously occur in the
field of view. To date, as a by-product of this survey, one sdBV star
has been discovered. It was assigned the designation
RAT\,J0455+1305 in accordance with the survey convention. Discovery observations along with 
Fourier analysis and a classification spectrum were published by \cite{ramsey06}. RAT0455
is 17.2\,mag in the the V filter and the amplitude spectrum calculated from the discovery
data revealed only one frequency in the ($p-$)mode region with a relatively high
amplitude.  Placing RAT0455 on the logT$_{\rm eff}$ and $\log g$ diagram (Fig.\ref{hr}), we can see that it is similar to other hybrid stars, and indeed \cite{baran10} detected peaks at low frequencies which were assigned as ($g-$)modes.

\begin{table}
 \centering
 \begin{minipage}{83mm}
  \caption{Results of the pre-whitening process for KPNO data. The phases are given at
mean epoch 2455145.89952. The numbers in parentheses are the errors of the last digits and the
residual noise level is 0.26\,mma.}
  \label{list}
  \begin{tabular}{@{}cccc@{}}
  \hline
mode                & Frequency [mHz]  & Amplitude [mma]  & phase [rad] \\
\hline
f$_{\rm A}$      &     0.22930(21)      &            2.2(2)           & 1.53(10) \\
f$_{\rm B}$      &     0.24910(36)      &            1.3(2)           & 5.99(17) \\
f$_{\rm C}$      &     0.27837(22)      &            2.1(2)           & 3.44(10) \\
f$_{\rm D}$      &     0.30793(25)      &           1.7(2)            & 1.52(12) \\
f$_{\rm E}$      &     0.31398(21)      &            2.1(2)           & 0.92(10) \\
f$_{\rm F}$      &     0.48251(37)      &            1.1(2)           & 2.68(19) \\
f$_{\rm 1}$      &     2.68044(2)        &           18.8(2)          & 0.90(1) \\
f$_{\rm 2}$      &     2.74899(10)      &             3.9(2)          & 4.25(5) \\
f$_{\rm 3}$      &     2.67673(9)        &             5.3(2)           & 5.45(4) \\
f$_{\rm 4}$      &     3.59305(33)      &             1.2(2)          & 5.47(17) \\
f$_{\rm 5}$      &     4.40179(36)      &             1.1(2)          & 0.61(18) \\
f$_{\rm 1+2}$ &     5.42943              &             0.8(2)          & 5.50(24) \\
\hline
\end{tabular}
\end{minipage}
\end{table}

\begin{figure}
\includegraphics[width=83mm]{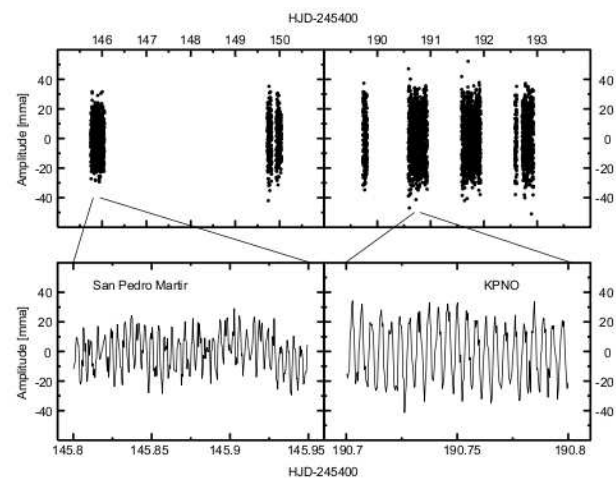}
\caption{Photometry obtained during late 2009 at two sites: San Pedro Martir (left panels)
and KPNO (right panels). Closer views are in the lower panels.}
\label{data}
\end{figure}

\begin{figure}
\includegraphics[width=83mm]{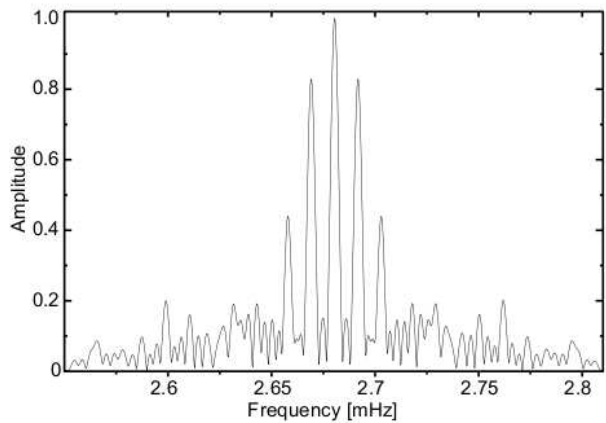}
\caption{Window function of data taken at KPNO. Large diurnal aliasing indicates the single-site nature 
of the data.}
\label{window}
\end{figure}

\begin{table}
 \centering
 \begin{minipage}{83mm}
  \caption{Frequency ratios of consecutive ($p-$)mode regions for RAT0455 and Bal09. 
The ratios were calculated using the highest amplitude frequencies
in each group. The fourth group in RAT0455 was not detected.}
  \label{ratio}
  \begin{tabular}{@{}ccc@{}}
  \hline
 groups  & RAT0455 & Bal09 \\
\hline
1/2          &      0.746     &   0.745 \\
2/3          &      0.816     &   0.813 \\
3/4          &      --             &   0.850 \\
\hline
\end{tabular}
\end{minipage}
\end{table}

\begin{figure}
\includegraphics[width=83mm]{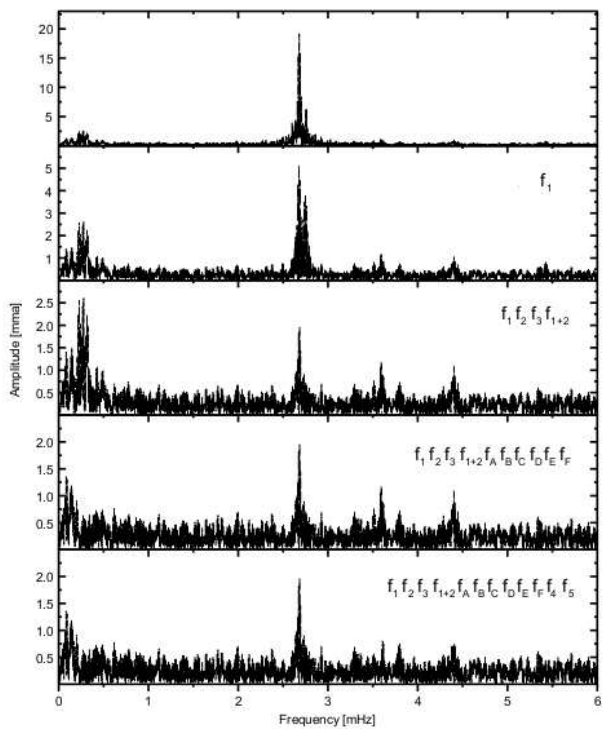}
\caption{Amplitude spectra for the original data and after some prewhitening steps for the KPNO data.
Each panel indicates the number of frequencies removed from the original spectrum.}
\label{spectrum}
\end{figure}

\begin{figure}
\includegraphics[width=83mm]{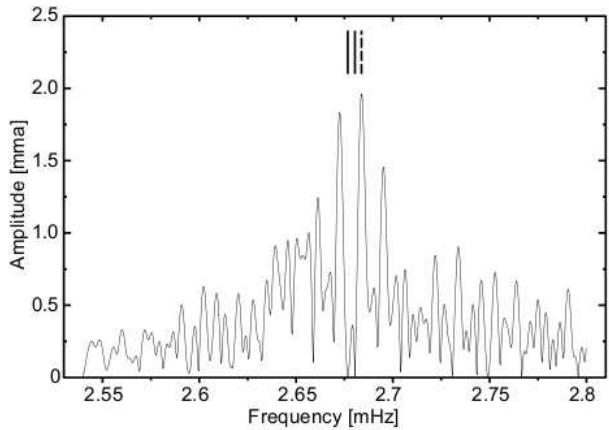}
\caption{Amplitude spectrum around the dominant peak. Location of two prewhitened
peaks are indicated by solid lines. The dashed line indicates the position of the
remaining peak and is added to indicate a potential triplet.}
\label{dominant}
\end{figure}

\section{Observations}
\subsection{First data}
Our first effort to obtain follow-up time-series photometry of RAT0455
was in January 2009 using the 1.5\,m telescope at San
Pedro Martir Observatory in Baja, Mexico. 
To maximize signal, the observations were conducted without a
filter. Photometric data were obtained during three of five
nights. Unluckily, non-photometric conditions deteriorated the
data quality. Those data confirmed 
the dominant periodicity at the frequency and amplitude reported in the discovery 
paper, and revealed three new small-amplitude peaks in the long period domain. The latter
are usually identified as ($g-$)modes in sdB stars. Detection of both short and long period
modes confirmed that RAT0455 is indeed a hybrid star and 
increased the number of known hybrid sdBV stars to 4. Details on observations, 
analysis and conclusions can be found in the paper by \cite{baran10}.

\subsection{Late 2009 data}
One of the conclusions of our previous paper was that
more data were needed to detect more, lower amplitude frequencies (if any existed) and 
to search for other features, such as
evenly spaced periods and frequencies or multiplets. Such features
may help to constrain pulsation parameters, in particular the three quantities that
describe the pulsation geometry and are usually denoted as $n$, $l$ and $m$.
In late 2009, we were allocated time on two telescopes: the 1.5\,m at San Pedro Martir and the 2.1\,m at KPNO. At the former we had five nights while at the latter we planned 7 nights. Again the weather was not cooperative and the overall duty cycle was only about 60\% of the available time. At San Pedro Martir we used a Marconi CCD with a square 3 by 3\,arcmin field of view. The exposure time was 30\,sec with 7\,sec read out giving a cadence of 37\,sec. As in January we did not use any filter. At KPNO we used an Apogee CCD also with a 3 by 3\,arcmin field of view. We used exposure time of 30\,sec, with a cadence of 32\,sec, but here we used a BG40 filter to improve the contrast with the sky background. The details are given in Table \ref{log}. All data were calibrated for instrumental effects and the brightness was extracted by means of PSF fitting and aperture photometry. Those data which were characterized by smaller scatter were taken for the final analysis. The data were corrected for color extinction with each night treated separately.

Because of non-photometric conditions on 11 Nov 2009, those data were very
noisy and not used in our Fourier analysis. They are not plotted in
Figure \ref{data}, which shows data obtained at San Pedro Martir (left panels)
and at KPNO (right panels).  The upper panels for each site show all data,
while the lower panels zoom in on a part of one night from each site.
From this plot one can see the different quality of data obtained at the two
different sites. Data from the 1.5\,m telescope are noisier, and the lack of filter
results in a broader bandpass which reduced the pulsation amplitudes.

\section{Amplitude spectrum}
To derive amplitudes, frequencies and phases of the pulsations a
Fourier analysis and pre-whitening process were employed. At each step of pre-whitening, 
sinusoidal terms were improved by means of a non-linear least-squares method. Since the observations in San Pedro 
Martir were performed without any filter, and were separated in time from the KPNO run, data from different sites 
were analyzed separately.

\subsection{San Pedro Martir data}
We only obtained useful data on two nights from San Pedro Martir and because of a flux discontinuity, 
part of 14 Nov data were removed. This results in a low duty cycle for these two nights of data, with an extra 
gap in the second night.
Combining these two nights causes an aliasing pattern of 1/(4\,days) =
2.89\,$\mu$Hz from any real peak which complicates the pre-whitening process.
Analysis of the combined data revealed a timing
issue in the data from 14 November. It was caused by incorrect times being
written to the FITS header. Unfortunately, the reason for this failure
remains unknown and so these data were not usable -- we are therefore left with just 
one night's data. As such, these data did not detect any new
frequencies. The only periodicity detected with S/N\,$>$\,4 is the dominant peak
at a frequency of 2.6782(7)\,mHz, an amplitude of 13.1(4)\,mma and a phase of
4.19(3)\,rad at epoch 2455145.89952. This poor estimation of the frequency may be caused by bad coverage during our last night.

\subsection{KPNO data}
\label{dec09}
In December 2009, observations were obtained at the 2.1\,m telescope at
KPNO. Of seven allocated nights, data were collected during the first four. In total, more than 27\,hrs of data were gathered. Because of the larger aperture at KPNO compared to San Pedro Martir, these data show a higher signal-to-noise, and even more importantly, no timing issues were present. Data from all four nights were combined and analyzed by means of Fourier analysis. Figure \ref{window} shows the window function calculated from these data. Frequencies, amplitudes and phases are 
presented in Table \ref{list}. Figure \ref{spectrum} shows residual amplitude spectra after a few steps in the 
pre-whitening process. The frequency resolution for these data is about 3.5\,$\mu$Hz. 

During these observations, we detected more frequencies than
\cite{baran10} did. Some were previously detected 
and are confirmed with these data while others are detected for the
first time. Unfailingly, it is easy to say that our dominant peak f$_1$ is the
same detected by \cite{baran10} and our f$_2$ is equivalent to theirs.
For other frequencies, it is not as easy to compare the two investigations. Peaks f$_4$ and
f$_5$ from our analysis were not detected before. Their locations are relatively regular and similar to
those found in Balloon 090100001 (Bal09) \citep{baran05}. 
In the $g$-mode region, it is difficult to compare peaks between seasons
because the spectrum is very dense, so we do not know if two peaks with similar 
but not identical frequencies are the same (to within the errors) or if they are completely different peaks.
Frequency proximity may suggest a similarity between f$_{\rm B}$ and f$_5$ or
f$_{\rm E}$ and f$_4$, where the former peak is detected in this analysis while the latter
in \citet{baran10}. This results in either 5 or 9 independent ($g-$)modes
detected so far.

Also, we cannot investigate amplitude stability as
these data were not obtained in the same filter as \cite{baran10},
so amplitudes cannot be directly compared.
Nevertheless, they are for the dominant peak, i.e. 18.8\,mma and
17.1\,mma (in the previous season).  Though the newest data were taken with
the BG40 filter, we can conclude that the amplitude has decreased. Amplitudes
are decreased when no filter is used as it is averaged
out through the wider bandpass.
In total we detected 5 high-frequency peaks indicative of $p$-modes
and 6 peaks in the low-frequency $g$-mode regime. We also detect one combination
frequency of the dominant mode and f$_2$. Its amplitude is very small and below
the S/N\,$>$\,4 criterion, but as a combination peak it is also included in
the solution.

What may be surprising is why, in the lowest panel in Figure \ref{spectrum}, the residual
amplitude spectrum still has a peak with relatively high amplitude which
satisfies the above criterion. Its frequency falls near 2.6838\,mHz and pre-whitening
of this peak was unsuccessful. Including this peak grossly degenerated the fits to other peaks in the region.
It should
be noted that f$_1$ is separated from f$_3$ by approximately the formal
frequency resolution, and this may cause unreliable fits via non-linear least-squares method. 
The same problem plagues the periodicity that we cannot pre-whiten. Its separation from the dominant peak 
is almost the same as for f$_3$. We suspect that with the current temporal
coverage we cannot fully resolve all of the peaks around the
dominant peak. As such, we emphasize that frequencies
with small amplitudes in the vicinity of f$_1$ must be considered with caution.

\section{Discussion -- Preliminary mode identification}
Using our detected frequencies, we would like to correlate them with
the three pulsation parameters which describe pulsation geometry, i.e.
$l$ and $m$, and the radial order $n$. Mode identification is essential for constructing 
theoretical models of stellar interiors. Experience from other
pulsating sdB stars show that reliable mode identifications are not easy.

These stars are faint, and that is why observations typically only use one filter.
In such cases, identifying features like multiplets can help to deduce one
or more pulsation parameters. Even narrowing the range of possible values
can be an advantage to provide tighter constraints on theoretical models. Unfortunately,
multiplets are not common among sdBV stars \citep{reed08}. Usually peaks do not appear in
a systematic way and then the only way to probe stellar models is to
test each frequency against reasonable values of various free parameters. A solution is then
attributed to the best fit between the observed frequencies (and global stellar parameters) and the
model values \citep{charp05}.

In the case of RAT0455, a dozen frequencies have been
detected. Luckily, in the amplitude spectrum there are features which can
help in mode identifications. They are similar to what was found in Bal09 \citep{baran05}.
First is the groupings of ($p-$)modes. These are not spread over
a wide range of frequencies but instead they are located in specific regions only.
In the case of Bal09 they were interpreted as consecutive overtones of the
radial modes. This means the adjacent groups contain modes which
differ in $n$ by 1. In RAT0455 such groupings are also present. However, since
the noise level is higher, we could detect only one peak in each of two groups
which satisfy the S/N\,$>$\,4 rule. In the residual amplitude spectrum (lowest panel in
Figure \ref{spectrum}) one can see a signal excess near to f$_4$ and f$_5$. It is
likely that those groups contain more peaks but with amplitudes too small to
be detected with our data. 

The frequency ratio between consecutive regions (calculated
using the highest amplitude frequency in each group) as well as those for Bal09
for comparison are presented in Table \ref{ratio}.
Our results are very similar to Bal09 which suggests 
that the interpretation for Bal09 may apply to RAT0455. Therefore,
the dominant peak could be a radial fundamental mode, while the second
and third groups of frequencies might contain consecutive radial overtones. 
This means that if f$_1$ is the radial fundamental ($n$\,=\,$l$\,=\,0) then f$_4$ is the first overtone 
($n$\,=\,1, $l$\,=\,0) and f$_5$ is the second overtone ($n$\,=\,2, $l$\,=\,0). Of course, we could make a mistake 
assigning the dominant frequency to the fundamental mode instead of a higher overtone with radial order of $n$. 
Then f$_4$ and f\,$_5$ would have $n$\,=\,$n$+1 and $n$\,=\,$n$+2, respectively. This includes an assumption
 that the highest amplitude frequency within each group is a radial ($l$\,=\,0) mode, as they suffer the least from surface cancellation effects. It does not have to be true, but mode identifications of other sdBV stars shows this to be the most likely case \citep[][]{baran08,telting04}.

Another feature even more useful for mode identifications is the existence of multiplets -- three or more peaks equally 
spaced in frequency as a result of rotation (similar to the Zeeman effect in atomic structure). With multiplets, 
we can almost instantly
assign $l$ and $m$ values to each component. In addition, a rotation
period can be derived from the splittings. However Table \ref{data} does not reveal any obvious multiplets.  
Figure 5 shows the  region surrounding the dominant
peak after removing f$_1$, f$_2$ and f$_3$.
It is obvious that there is another peak satisfying S/N\,$>$\,4 rule,
however its removal was not possible as it leads to spurious solutions (see \ref{dec09}). It is
possible evidence for a multiplet, but the insufficient frequency resolution
does not allow us to resolve such closely spaced peaks.
The two identified (and removed) peaks along
with the one in the residual amplitude spectrum may form an almost equidistant
triplet with a mean splitting of $\Delta$f\,=\,3.5\,$\mu$Hz. If we assume that it is a rotationally split $l$\,=1 mode 
then the rotation period is $P_{\rm rot}=(1-C_{n,l})/\Delta f$. Assuming that the Ledoux constant (C$_{n,l}$) for $l$\,=\,1 
is small \citep[$<$0.05][]{charp00}, we derive $P_{\rm rot}$ to be about 3.3\,days. Of course if these peaks are 
components of a triplet, then the dominant peak must have a non-zero value of $l$ (i.e. $l$\,=\,1).  If so, this is 
contrary to the conclusion from the previous section that this peak may be radial. Additionally, it would mean that no
radial modes are observed, despite their lack of any geometric cancellation. Another possibility is these
three peaks do 
not form a triplet, but rather only two of them i.e. f$_3$ and the
residual peak are parts of a triplet with a third member below our detection limit. The splitting would
then be around 7\,$\mu$Hz and the corresponding rotation period close to 1.65\,days.

\section{Conclusions}
We presented analysis of photometric data for one of the
faintest sdBV stars, RAT0455. Despite its faintness, we obtained sufficient
data to detect more, previously undetected frequencies and  confirm the hybrid nature of RAT0455.
Similar to Bal09, we find non-uniform groupings of short-period $p$-modes and like  Bal09, assign them to be
the radial fundamental mode and two overtones. Such mode constraints apply directly to stellar models through
which we learn about stellar interiors.
A possible triplet is found, but higher frequency resolution is needed to confirm its existence.
Our results do not, at this point, provide high confidence in our mode identification ,
but this star looks very promising. Longer data
coverage with a lower noise level in the amplitude spectrum would likely detect additional pulsation frequencies
and answer these questions. That would make RAT0455 one of the very few sdBV stars with observationally constrained
mode identifications.

\section*{Acknowledgments}

AB gratefully appreciates funding from Polish Ministry of Science and Higher Education under project No. 554/MOB/2009/0. JTG was supported by the Missouri Space Grant Consortium, funded by NASA. LFM acknowledges financial support from the UNAM via PAPIIT grant IN114309.

\label{lastpage}

\end{document}